\pdfoutput=1
\documentclass[a4paper,11pt]{article}

\usepackage{jheppub} 

\usepackage[T1]{fontenc} 
\usepackage{lmodern}
\usepackage[utf8]{inputenc}
\usepackage[usenames,dvipsnames]{xcolor}
\usepackage{amsmath,amssymb,amsthm,mathtools,mathrsfs}

\usepackage[some]{background}
\usepackage{braket}
\usepackage{slashed}
\usepackage{graphicx}
\graphicspath{{Images/}}
\usepackage{bm}
\usepackage{tocloft}
\usepackage[bottom]{footmisc}
\usepackage{multirow}
\usepackage{hyperref}
\hypersetup{
    colorlinks=true,
    linkcolor=red,
    citecolor=blue,
    urlcolor=red,
    linktocpage,
}
\usepackage{subfig}
\usepackage{comment}

\newcommand\q{{\mathbf{q}}}
\newcommand\p{\mathbf{p}}
\newcommand\x{\mathbf{x}}

\theoremstyle{definition}

\theoremstyle{definition}

\theoremstyle{definition}

\theoremstyle{plain}

\theoremstyle{remark}
\newtheorem*{rem*}{\protect\remarkname}
\theoremstyle{remark}

\theoremstyle{plain}

\theoremstyle{plain}

\theoremstyle{plain}

\providecommand{\definitionname}{Definition}
\providecommand{\examplename}{Example}
\providecommand{\exercisename}{Exercise}
\providecommand{\propositionname}{Proposition}
\providecommand{\remarkname}{Remark}
\providecommand{\theoremname}{Theorem}

\title{Rotons in Anyon Superfluids}

\author{Yi-Hsien Du, Umang Mehta, Dam Thanh Son}
\affiliation{Kadanoff Center for Theoretical Physics, University of Chicago}
\emailAdd{yhdu@uchicago.edu}
\emailAdd{umangmehta@uchicago.edu}
\emailAdd{dtson@uchicago.edu}

\abstract{We consider the problem of calculating the excitation
  spectrum of a gas of nonrelativistic anyons.  When the anyons have
  statistics close to fermionic and the statistical angle has the form
  $\theta=\pi (1-\frac 1k)$ where $k$ is a large integer, the problem
  can be solved by employing the method of bosonization, which maps
  the problem to that of an infinite number of bosonic excitations
  coupled to a $U(1)$ Chern-Simons gauge field.  The spectrum
  consists of a Goldstone boson branch and a large number of
  massive branches, each having roton minima and maxima.  The
  dispersion curves asymptote to the Landau levels at large
  momentum.}

\begin{document} 

\maketitle

\section{Introduction}\label{int}

It was recognized a long time ago~ \cite{Leinaas1977} that in two
spatial dimensions quantum mechanics allows a much richer set of
statistics besides Bose or Fermi statistics.  The simplest
nontrivial possibility is the abelian anyonic statistics, in which the
wave function changes by a complex phase $e^{i\theta}$ different from
1 or $-1$ under the exchange of two particles.  Such statistics appears
in many quantum Hall states, for example in the $\nu=1/3$ Laughlin
state~\cite{Laughlin:1983fy}, as a property of the quasiparticle
excitations~\cite{Arovas:1984qr,Manfra2020}.

The statistics of an ensemble of anyons presents a nontrivial
problem~\cite{Arovas:1985yb}.  When the statistical phase $e^{i\theta}$
is different from 1 or $-1$, the statistical interaction between the
anyons cannot be ignored: a gas of anyons is never a
``non-interacting'' gas even when no additional interaction between
the anyons is present.
One particular case is better understood than the others: when the
statistical phase has values $\theta=\pi(1-1/k)$ where $k$ is an
integer.  This includes the case of the so called
``semions''---particles with statistics right in between that of the
boson and the fermion ($k=2$).  It can be shown (see section \ref{eft_as} below)
that the system is a superfluid at large $k$, and perhaps for all
integer values of $k$ down to $k=2$.  This possibility was briefly
considered, but then discarded, as a candidate for high-$T_c$
superconductivity.

At lowest energies the excitation spectrum of an anyon gas contains
only the Nambu-Goldstone boson of superfluidity.  In this paper we
investigate the spectrum of excitations of this anyon gas beyond the
Nambu-Goldstone boson.  We show that at large $k$ there are two
momentum scales in the problem: the Fermi momentum $p_F$ and a smaller
momentum scale $p_F/k$.  The theory containing one Nambu-Goldstone
boson describes only the physics below the momentum scale $p_F/k$ (and the
corresponding energy scale).  We develop an effective theory that is
valid over a much larger energy and momentum scale: the
momentum cutoff of this theory is $p_F$.  We find that the
gas exhibits a number of branches, each containing roton minima and
maxima.  The number of branches and the number of roton minima and
maxima tends to infinity as $k\to\infty$.  Such roton-like features
have been seen in previous calculations using the random-phase
approximation of the ``semion'' ($k=2$) gas \cite{Yang1991,Cheng1993}. While our calculations cannot be extrapolated to this case since we require the coupling $1/k$ to be small, our
result is more reliable precisely due to the small parameter.

The paper is organized as follows. In section \ref{eft_as} we review
the effective theory
for the Nambu-Goldstone boson. In section \ref{bos} we develop a bosonized description for Fermi surfaces coupled to dynamical gauge fields in a Hamiltonian formalism, using the `higher-spin' theory of \cite{hs}, and use it to compute the equations of motion for the Fermi surface. In section \ref{eom_sol} we solve the equations of motion to obtain a transcendental equation whose solutions are the dispersion relations for the various modes of the Fermi surface. In section \ref{spect} we analyze the resulting spectrum of excitations and present the relation to Landau levels through small and large momentum asymptotics. In section \ref{kct} we develop the effective theory further to eliminate an unphysical zero mode from the spectrum by generalizing Kelvin's circulation theorem to our theory.

\section{Effective theory for the Nambu-Goldstone boson of anyon superfluids}\label{eft_as}

We consider a gas of nonrelativistic anyons which have a statistical
angle $\theta$ of the form
\begin{equation}
    \theta = \pi \left( 1 - \frac{1}{k} \right).
\end{equation}
where $k$ is an integer.  The case $k=1$ corresponds to bosons and
$k=2$ to the so-called semions.  We will consider the limit of large
$k$, for which reliable calculations are possible.  This anyon gas can
be modeled as a system of non-relativistic fermions at non-zero
density coupled to a $U(1)_{-k}$ Chern-Simons theory, with the action
\begin{equation}\label{ac}
  S = \int d^3 x \left( i\psi^\dagger  D_0  \psi - \frac{1}{2m} D_i \psi^\dagger D_i \psi - \frac{k}{4\pi} ada \right) = S_\psi - S_{\text{CS}_k},
\end{equation}
where $D_\mu = \partial_\mu - ia_\mu$ is the covariant derivative, and
we have denoted by $S_\psi$ and $S_{\text{CS}_k}$ the fermionic and
Chern-Simons parts of the action, respectively. The non-zero density of
fermions fixes the vacuum expectation value (VEV) of the magnetic
field $\langle b \rangle \ne 0$ via the $a_0$ equation of motion
\begin{equation}\label{density}
	\bar{\rho} \equiv \langle \psi^\dagger \psi \rangle = \frac{k}{2\pi} \langle b \rangle.
\end{equation}

As far as we know, there has not yet been any experimental realization of this model, since in all known experimental systems with abelian anyonic excitations, such as fractional quantum Hall states, the anyons carry non-vanishing electric charge, while in this model they are neutral.

Using the random phase approximation (RPA), Refs.\ \cite{Fetter1989,asc}
show that the
anyon gas forms a superfluid  with a
linearly dispersing Nambu-Goldstone boson.
The most illuminating way to demonstrate this fact is to note that
integrating out the fermion $\psi$ cancels the Chern-Simons term and
results in the Maxwell action for $a_\mu$~\cite{Banks1990}.  The
Nambu-Goldstone boson is just the massless photon in the resulting
(2+1)D quantum electrodynamics.

Let us recall in more detail how this happens.  At large $k$, we split
up the gauge field into mean-field and fluctuating components $a =
\bar{a} + \delta a$.  At the mean-field level, \eqref{density} implies
that the fermions fill exactly $k$ Landau levels, resulting in a
gapped system with the gap given by the cyclotron frequency $\omega_c
= b/m$.  For the physics below the $\omega_c$ energy, the fermions can
then be integrated out to obtain the low energy effective action for
the fluctuation in the gauge field.

Integrating out the fermions in $k$ filled Landau levels amounts to
computing the generating functional for the electromagnetic response
of the integer quantum Hall effect.  That can be done using the
formulas derived in Ref.~\cite{nguyen_gromov} (see Eq.~(66) therein)
to quadratic level and to lowest order in the derivative expansion the
result reads
\begin{equation}
	\begin{split}
		e^{iW[a]} &= \int \mathcal{D}\psi ~ e^{i S_\psi},\\
		W[a] &= \frac{k}{4\pi} \int \left( ada + \frac{m}{b^2}e^2 - \frac{k}{m} (\delta b)^2
                \right) + \mathcal{O}(\partial^4).
	\end{split}
\end{equation}
In the anyon superfluid case, the low energy effective action is then given by the following,
\begin{equation}
	\begin{split}
		e^{iS_\text{eff}[\delta a]} &= \int [\mathcal{D}\psi] e^{iS} = \left( \int [\mathcal{D}\psi] e^{iS_\psi} \right) e^{-iS_{\text{CS}_k}},\\
		&= e^{i(W - S_{\text{CS}_k})},\\
		\implies S_\text{eff}[\delta a] &= \frac{k}{4\pi} \int \left( \frac{m}{b} e^2 - \frac{k}{m} (\delta b)^2
                \right) + \mathcal{O}(\partial^4),
	\end{split}
\end{equation}
This action has a gapless excitation, since the lowest order term is
the Maxwell term with speed $c_s^2 = kb/m^2$. A gapless gauge field in
$2+1$ dimensions is dual to a compact scalar and hence describes a
superfluid.

Note that this effective theory captures only the dynamics of the
Nambu-Goldstone boson of the anyon superfluid at low momenta.
The momentum and energy cut-offs for this
theory are given by the cyclotron frequency $\omega_c = b / m$, or
equivalently by the inverse magnetic length $1/l_b = \sqrt{b}$. In
order to explore the spectrum beyond these cut-offs, we need to employ
a different approach.

\section{Coupling Fermi surfaces to gauge fields via bosonization }\label{bos}

In the large $k$ regime where the theory is weakly coupled, there
exists an intermediate regime of momentum above the cutoff momentum of
the theory of the Nambu-Goldstone boson, but below the Fermi momentum
$p_F$.  This is the regime we will now be interested in.
One can obtain a quantitative
description of the system by bosonizing the Fermi surface using the
`higher-spin' formalism, developed in Refs.~\cite{hs,fqh} for
fractional quantum Hall states near half filling, and coupling it to
dynamical fluctuations in the gauge field around the uniform magnetic
field.

\subsection{Fermi surface bosonization in a background magnetic field}\label{hs_rev}

We begin by reviewing the bosonization procedure for a Fermi surface.
First we neglect the fluctuations of the gauge field, considering the
latter as a uniform static background.  The Fermi surface is
bosonized by treating its shape as a dynamical field, decomposed into
angular channels,
\begin{equation}
    p_F(x,\theta) = p_F^0 + \sum_{n=-\infty}^\infty u_n(x,\theta) e^{in\theta},
\end{equation}
where $x=(t,\mathbf{x})$ is a spacetime coordinate, and $\theta$ is
the angular coordinate in momentum space. We will drop the superscript
0 on $p_F^0$ from here on for convenience. The commutations relations
for the Fourier components $u_n$ can be obtained from single-particle
Poisson brackets in phase space \cite{hs},
\begin{equation}\label{hsalg}
    [u_m(\mathbf{q}), u_n(\mathbf{q}')] = \frac{2\pi}{p_F} \left( -\frac{bm}{p_F} \delta_{m+n,0} + q_z \delta_{m+n,1} + q_{\bar{z}} \delta_{m+n,-1} \right) (2\pi)^2 \delta(\mathbf{q}+\mathbf{q}') + \mathcal{O}(u),
\end{equation}
where $q_z = (q_1 - iq_2)/2$.

Let us recall one way the commutation relation~(\ref{hsalg}) can be
derived (see also Refs.~\cite{fradkin_bos, haldane_bos, ersatz_fl}).
Denote by $n_\mathbf{p}(\mathbf{x})$ the distribution function which
is 1 when $\p$ is located inside the local Fermi surface at point $\x$
and 0 when $\p$ is outside. Using two arbitrary functions $f$, $g$ in
phase space, we can construct two operators
\begin{equation}
	\begin{split}
		F &= \int \frac{d^2x ~ d^2p}{(2\pi)^2} f(\mathbf{x},\mathbf{p}) n_\mathbf{p}(\mathbf{x}),\\
		G &= \int \frac{d^2x ~ d^2p}{(2\pi)^2} g(\mathbf{x},\mathbf{p}) n_\mathbf{p}(\mathbf{x}).
	\end{split}
\end{equation}
which are implicitly functionals of the shape of the Fermi surface.
The commutator of these two operators must then obey
\begin{equation}\label{arb_comm}
  [F, \, G] = -i\!\int \frac{d^2x ~ d^2p}{(2\pi)^2} \{ f , \, g \} (\mathbf{x},\mathbf{p}) n_\mathbf{p}(\mathbf{x}),
\end{equation}
where
\begin{equation}
	\{ f , g \} = \frac{\partial f}{\partial p_i} \frac{\partial g}{\partial x_i} - \frac{\partial f}{\partial x_i} \frac{\partial g}{\partial p_i} - b \epsilon^{ij} \frac{\partial f}{\partial p_i} \frac{\partial g}{\partial p_j},
\end{equation}
is the canonical Poisson bracket of two phase space functions in the
presence of a magnetic field. For a zero temperature Fermi surface,
$n_\mathbf{p}(\mathbf{x})$ is completely specified by the shape of
the Fermi surface, and we can write $n_\mathbf{p}(\mathbf{x}) =
\theta(p_F + u(\mathbf{x},\theta) - |\mathbf{p}|)$ as a step function
supported within the surface. Evaluating $F, G$ to linear order in $u$
results in an expression for the left-hand side of \eqref{arb_comm} in
terms of the commutator $[u(\mathbf{x},\theta),
  u(\mathbf{x}',\theta')]$, which can be equated to the expression
obtained by evaluating the right-hand side of \eqref{arb_comm} to
zeroth order in $u$. This results in an expression for the commutator
of $u$ with itself, which can be Fourier transformed to obtained
\eqref{hsalg}.

The kinetic equation of Landau's Fermi liquid theory is equivalent
to the equations of motion derived from the quadratic
Hamiltonian
\begin{equation}
  H = \frac{v_F p_F}{4\pi} \int\! d^2x \sum_{n=-\infty}^{\infty}
  (1+F_n) u_n(\mathbf{x}) u_{-n}(\mathbf{x}).
\end{equation}
and the commutation relations~(\ref{hsalg}).  Here $p_F$ and $v_F$ are
the Fermi momentum and Fermi velocity, respectively, and $F_n$ are the
Landau parameters.  In the case of a weakly-interacting anyon gas at
large $k$ the Landau parameters are expected to be small, while $p_F$
and $v_F$ are related to the magnetic field VEV $b$ by
\begin{equation}
    \frac{p_F^2}{4\pi} = \frac{kb}{2\pi}, \qquad v_F^2 = \frac{p_F^2}{m^2} = \frac{2kb}{m^2}.
\end{equation}

In Fourier space the equations of motion has the form of a recursion
relation\footnote{Our convention for Fourier transform is $u_n(x) =
\int_q u_n(q) e^{iq_\mu x^\mu}$ with the mostly positive signature.},
\begin{equation}\label{eom}
	\left( \omega + \frac{b}{m} n \right) u_n = v_F \left( q_z u_{n-1} + q_{\bar{z}} u_{n+1} \right).
\end{equation}
We will also need the formulas relating the charge density and current to the
the phase space distribution:
\begin{equation}\label{curr}
    \begin{split}
        \rho &= \int_\mathbf{q} n_\mathbf{q}(x) = \bar{\rho} + \frac{p_F}{2\pi} u_0,\\
        j^i &= \int_\mathbf{q} n_\mathbf{q}(x) \frac{q^i}{m} = \frac{p_F^2}{4\pi m} \begin{pmatrix} u_1 + u_{-1}\\ i(u_1 - u_{-1}) \end{pmatrix}.
    \end{split}
\end{equation}
Here $\int_\q\equiv\int\!(d^2q)/(2\pi)^2$, and $\bar{\rho} = p_F^2 /
4\pi$ is the average fermion density. These equations give a physical
interpretation to the Fourier components $u_0, u_{\pm 1}$---they are
the particle number and momentum density to linear order.

The natural cutoffs for the description that has just been presented are
the Fermi momentum $p_F$ and the Fermi energy.

\subsection{Gauge field fluctuations}\label{gf_fluc}

The formalism that has been presented needs to be modified to take
into account fluctuating gauge fields.  In the Hamiltonian formalism
the spatial components $\delta a_i$ are dynamical variables; in
contrast the temporal component $a_0$ is a Lagrange multiplier
imposing a constraint.  We need to find the commutators involving
$a_i$.

From the Chern-Simons action one reads out the commutators
between the fluctuations of $a_1$ and $a_2$:
\begin{equation}\label{aa-comm}
  [\delta a_z(\q), \delta a_{\bar{z}}(\q')]
  = \frac{\pi}{k} (2\pi)^2 \delta(\q+\q'),
\end{equation}

In order to obtain the modified Poisson brackets of the Fermi surface
Fourier components, one can make the following observation.  Since the
fields $u_0, u_{\pm 1}$ constitute the current and density, they can
be identified with the fermion current and density obtained from the
microscopic action:
\begin{equation}
	\begin{split}	
		\rho &= \frac{\delta S_\psi}{\delta a_0} = \psi^\dagger \psi,\\
		j^i &= \frac{\delta S_\psi}{\delta a_i} = -\frac{i}{2m} \psi^\dagger \overleftrightarrow{\mathcal{D}}_i \psi.
	\end{split}
\end{equation}
Comparing with \eqref{curr}, we can identify $u_0$ and $u_{\pm 1}$ with
the microscopic operators
\begin{equation}
	\begin{split}
		u_0 &= \frac{2\pi}{p_F} \left( \psi^\dagger \psi - \bar{\rho} \right),\\
		u_1 &= -\frac{2\pi i}{p_F^2} \psi^\dagger \overleftrightarrow{\partial}_z \psi - \frac{4\pi}{p_F^2} \psi^\dagger \psi a_z,\\
		u_{-1} &= -\frac{2\pi i}{p_F^2} \psi^\dagger \overleftrightarrow{\partial}_{\bar{z}} \psi - \frac{4\pi}{p_F^2} \psi^\dagger \psi a_{\bar{z}}.
	\end{split}
\end{equation}
From these expressions and Eq.~(\ref{aa-comm}), we find that $u_1$ and
$u_{-1}$ have nonzero commutators with $a_{\bar z}$ and $a_z$.  It is
convenient to separate the parts corresponding to the fluctuations of
$a_i$ from the operators $u_{\pm1}$.
Writing $a_i = \bar{a}_i + \delta a_i$, where $\delta a_i$ is the fluctuation of the gauge field around the configuration $\bar{a}_i$ that generates the uniform magnetic field, as well as using the fact that $\psi^\dagger \psi = \bar{\rho} + \delta \rho = (p_F + u_0)^2/4\pi \approx p_F^2/4\pi + p_f u_0 /2\pi$, the above expressions can be linearized in the fields $(u_n, \delta a_i)$ to obtain
\begin{equation}
	\begin{split}
		u_1 &= \bar{u}_1 - \delta a_z,\\
		u_{-1} &= \bar{u}_{-1} - \delta a_{\bar{z}},\\
		\bar{u}_1 &= -\frac{2\pi i}{p_F^2} \psi^\dagger \overleftrightarrow{\partial_z} \psi - \frac{4\pi}{p_F^2} \psi^\dagger \psi \bar{a}_z,\\
		\bar{u}_{-1} &= -\frac{2\pi i}{p_F^2} \psi^\dagger \overleftrightarrow{\partial_{\bar{z}}} \psi - \frac{4\pi}{p_F^2} \psi^\dagger \psi \bar{a}_{\bar{z}},
	\end{split}
\end{equation}
where $\bar{u}_{\pm 1}$ is the fermion momentum density in the absence
of the fluctuations of the gauge field. Since the gauge-field
fluctuations do not modify any of the other fields (in particular, due
to the mismatch in spin) so
\begin{equation}
	\bar{u}_n = u_n, \qquad n \ne \pm 1.
\end{equation}
Since the only consequence of turning on gauge field fluctuations is to modify the microscopic definition of the fields $u_{\pm 1}$, we propose the following Poisson brackets for the fields:
\begin{subequations}\label{pb}
\begin{align}
	[\bar{u}_m(q), \bar{u}_n(q')] &= \frac{2\pi}{p_F} \left( -\frac{bm}{p_F} \delta_{m+n,0} + q_z \delta_{m+n,1} + q_{\bar{z}} \delta_{m+n,-1} \right) (2\pi)^2 \delta(q+q'), \label{pb-1}\\
	[\delta a_z(q), \delta a_{\bar{z}}(q')] &= \frac{\pi}{k} (2\pi)^2 \delta(q+q'),\\
	[\bar{u}_n(q), \delta a_i(q')] &= 0,
\end{align}
\end{subequations}
where the second line is obtained from the Chern-Simons action for the
gauge field. Note that $b$ on the right
hand side of Eq.~(\ref{pb-1}) is the VEV, and does not include
fluctuations. The reason we don't include fluctuations is because we
are working at linear order in $u_n, \delta a_i$ at the level of
equations of motion. We also note, for future use, the commutators,
\begin{equation}
	\begin{split}
		[u_1(q), \delta a_{\bar{z}}(q')] &= -\frac{\pi}{k} (2\pi)^2 \delta(q+q'),\\
		[u_{-1}(q), \delta a_z(q')] &= \frac{\pi}{k} (2\pi)^2 \delta(q+q'),\\
		[u_1(q), u_{-1}(q')] &= 0,\\
		[u_1(q),\delta b(q')] &= \frac{2\pi}{k} q_z (2\pi)^2 \delta(q+q'),\\
		[u_{-1}(q),\delta b(q')] &= \frac{2\pi}{k} q_{\bar{z}} (2\pi)^2 \delta(q+q'),\\
	\end{split}
\end{equation}
where to obtain the last two lines we have used the fact that, in momentum space,
\begin{equation}
	\delta b (q) = 2 ( q_z \delta a_{\bar{z}} - q_{\bar{z}} \delta a_z ).
\end{equation}

As a consistency check one can verify that $[u_1, u_{-1}] = 0$.  This
has to be true since $u_{\pm 1}$ are the momentum densities of the
anyons, their spatial integrals are the generators of
translations. Since the anyons don't see any net magnetic flux due to
the fact that the flux attaches to the fermions, the generators of
translations must commute, as they do.

The Hamiltonian retains its usual form in terms of the un-barred fields:
\begin{equation}\label{ham}
	\begin{split}
		H = \frac{v_F p_F}{4\pi} \int &\frac{d^2 q}{(2\pi)^2} \sum_{n=-\infty}^\infty u_n(q) u_{-n}(-q)\\
		= \frac{v_F p_F}{4 \pi} \int &\frac{d^2 q}{(2\pi)^2} \Bigg[ \sum_{n=-\infty}^\infty \bar{u}_n(q) \bar{u}_{-n}(-q) - \left[ \bar{u}_1(q) \delta a_{\bar{z}}(-q) + \bar{u}_1(-q) \delta a_{\bar{z}}(q) \right]\\
		&- \left[ \delta a_z(q) \bar{u}_{-1}(-q) + \delta a_z(-q) \bar{u}_{-1}(q) \right] + \left[ \delta a_z(q) \delta a_{\bar{z}}(-q) + \delta a_z(-q) \delta a_{\bar{z}}(q) \right] \Bigg].
	\end{split}
\end{equation}

We can now commute the barred fields with the Hamiltonian to obtain the equations of motion for $\bar{u}_n$. These are the EFT analogs of the fermion equations of motion $\delta S / \delta \psi = 0$. We begin by commuting $\bar{u}_n$ with $H$ for $n\ne \pm 1$, using the first line of \eqref{ham}. Noting that
\begin{equation}
	[\bar{u}_n(q),u_m(q')] = [\bar{u}_n(q),\bar{u}_m(q')], \qquad n \ne \pm 1, m \in \mathbb{Z},
\end{equation}
simplifies that calculation and we find
\begin{equation}
	\dot{u}_n = \dot{\bar{u}}_n = -i \left( -\frac{b}{m} n u_n + v_F \left( q_z u_{n-1} + q_{\bar{z}} u_{n+1} \right) \right).
\end{equation}
Next, we compute the commutator of $\bar{u}_{\pm 1}$ with $H$ to obtain
\begin{equation}
	\begin{split}
		[\bar{u}_1(q), H] &= -\frac{b}{m} u_1 + v_F[q_z u_0 + q_{\bar{z}} u_2] + \frac{b}{m} \delta a_z(q),\\
		[\bar{u}_{-1}(q), H] &= \frac{b}{m} u_1 + v_F[q_z u_{-2} + q_{\bar{z}} u_0] + \frac{b}{m} \delta a_{\bar{z}}(q),
	\end{split}
\end{equation}
which implies the following equations for their time derivatives:
\begin{equation}
	\begin{split}
		\dot{\bar{u}}_1(q) &= -i \left( -\frac{b}{m} \bar{u}_1 + v_F[q_z u_0 + q_{\bar{z}} u_2] \right) + \frac{b}{m} \delta a_z(q),\\
		\dot{\bar{u}}_{-1}(q) &= -i \left( \frac{b}{m} \bar{u}_1 + v_F[q_z u_{-2} + q_{\bar{z}} u_0] \right) + \frac{b}{m} \delta a_{\bar{z}}(q).
	\end{split}
\end{equation}
Note that these can equivalently be written as equations for the time derivatives of the un-barred fields $u_{\pm 1}$:
\begin{equation}
	\begin{split}
		\dot{u}_1 &= -i \left( -\frac{b}{m} u_1 + v_F[q_z u_0 + q_{\bar{z}} u_2] \right) - \dot{\delta a}_z,\\
		\dot{u}_{-1} &= -i \left( \frac{b}{m} u_{-1} + v_F[q_z u_{-2} + q_{\bar{z}} u_0] \right) - \dot{\delta a}_{\bar{z}}.
	\end{split}
\end{equation}
Putting these together in a compact form and Fourier transforming to frequency space, we have
\begin{equation}\label{eomf}
	\left( \omega + \frac{b}{m} n \right) u_n  = v_F \left( q_z u_{n-1} + q_{\bar{z}} u_{n+1} \right) + i e_z \delta_{n,1} + i e_{\bar{z}} \delta_{n,-1},
\end{equation}
where we have used Weyl gauge $a_0 = 0$ to write $\dot{\delta a_i}$ in terms of the electric field $e_i = f_{i0}$.

The equations of motion for the gauge field can be obtained by equating the low energy expressions for the current to the variation of the Chern-Simons action,
\begin{equation}
    \begin{split}
		\frac{\delta S}{\delta a_\mu} &= \frac{\delta S_\psi}{\delta a_\mu} - \frac{\delta S_{\text{CS}_k}}{\delta a_\mu} = 0,\\
    	\implies j^\mu &= \frac{\delta S_{\text{CS}_k}}{\partial a_\mu} = \frac{k}{4\pi} \epsilon^{\mu\nu\rho} f_{\nu\rho}.
    \end{split}
\end{equation}
Plugging in the expressions \eqref{curr}, we find
\begin{equation}\label{eomg}
    \begin{split}
        \frac{k}{2\pi} \delta b &= \frac{p_F}{2\pi} u_0,\\
        ik e_z &= \frac{p_F^2}{2 m} u_1,\\
        - ik e_{\bar{z}} &= \frac{p_F^2}{2 m} u_{-1},
    \end{split}
\end{equation}
where $\delta b$ is the fluctuation of the magnetic field around its expectation value. Only the last two of these are genuine equations of motion in our Hamiltonian description, since they describe the time derivatives of $\delta a_i$. The first one is the Gauss law constraint and must be supplemented with our equations of motion. Naturally, our Poisson brackets and Hamiltonian must reproduce the second and third line of \eqref{eomg} as well. Commuting $\delta a_i$ with $H$, we find the expressions
\begin{equation}
	\begin{split}
		[\delta a_z(q), H] &= -\frac{b}{m} \left( \bar{u}_1(q) - \delta a_z(q) \right),\\
		[\delta a_{\bar{z}}(q), H] &= \frac{b}{m} \left( \bar{u}_{-1}(q) - \delta a_{\bar{z}}(q) \right),
	\end{split}
\end{equation}
which implies that the time derivatives are given by
\begin{equation}
	\begin{split}
		\dot{\delta a}_z(q) &= i\frac{b}{m} \left( \bar{u}_1(q) - \delta a_z(q) \right) = i \frac{b}{m} u_1(q),\\
		\dot{\delta a}_{\bar{z}}(q) &= -i \frac{b}{m} \left( \bar{u}_{-1}(q) - \delta a_{\bar{z}}(q) \right) = -i \frac{b}{m} u_{-1}(q),
	\end{split}
\end{equation}
in agreement with \eqref{eomg}.

Finally we have the primary constraint,
\begin{equation}
	C(q) \equiv p_F u_0(q) - k \delta b(q) \approx 0,
\end{equation}
where we use $\approx$ as the symbol for `weak equality' \`a la Dirac, i.e., the expression should be set to zero \textit{after} computing all Poisson brackets. $C$ commutes with all the un-barred fields, but doesn't commute with $\delta a_i$:
\begin{equation}
	\begin{split}
		[C(q), u_m(q')] &= 0\\
		[C(q), \delta a_z(q')] &= - \frac{2\pi}{k} q_z (2\pi)^2 \delta(q+q')\\
		[C(q), \delta a_{\bar{z}}(q')] &= - \frac{2\pi}{k} q_{\bar{z}} (2\pi)^2 \delta(q+q')
	\end{split}
\end{equation}
so it isn't trivially satisfied and does indeed constrain the phase space. What needs to be checked, in particular, is whether we generate any secondary constraints upon imposing
\begin{equation}
	i\dot{C}(q) = [C(q), H] = 0.
\end{equation}
The commutator with the Hamiltonian identically vanishes, since $C$ commutes with $u_m$ for every value of $m$, and hence we generate no new secondary constraints.

\section{Solving the equations of motion}\label{eom_sol}

We can use rotational symmetry to set, without loss of generality, $q_z = q_{\bar{z}} = q/2$. Furthermore, we introduce the dimensionless frequency $\tilde{\omega} = \omega/\omega_c = m \omega / b$, where $\omega_c$ is the cyclotron frequency, the dimensionless momentum $z = m v_F q / b = p_F q / b$, and $\lambda = 2k m / p_F^2 = m/b$ to simplify our equations of motion to
\begin{equation}\label{dimless-recurs}
    \begin{split}
      (\tilde{\omega} + n) u_n &= \frac{z}{2} ( u_{n-1} + u_{n+1})
      + i\frac{m}{b} e_z \delta_{n,1} + i \frac{m}{b} e_{\bar{z}} \delta_{n,-1},\\
        &u_1 = i\lambda e_z, \qquad \qquad \qquad u_{-1} = -i\lambda e_{\bar{z}}.
    \end{split}
\end{equation}
The Gauss law constraint doesn't play a role in solving these, since all it does it determine the fluctuation in the magnetic field $\delta b$ in terms of $u_0$, once $u_0$ has been solved for from these equations.

Imposing the boundary condition $|u_n| \rightarrow 0$ as $n\rightarrow \pm \infty$, we can solve the recursion relation \eqref{dimless-recurs} for $|n|\ge2$ and express all $u_n$ with $n\neq0$ through two functions $F(\tilde\omega, z)$ and $G(\tilde\omega,z)$:
\begin{equation}\label{sol1}
	u_n = \begin{cases}
		J_{n+\tilde{\omega}} F(\tilde{\omega},z), & n\ge 1,\\
		(-1)^n J_{-n-\tilde{\omega}}(z)G(\tilde{\omega},z), \quad & n\le -1.
    \end{cases}
\end{equation}
The remaining three equations of motion, corresponding to $n = 0, \pm 1$ in \eqref{dimless-recurs}, are
\begin{equation}
\begin{split}
   \tilde{\omega} u_0 &= \frac{z}{2}[J_{1+\tilde\omega}(z) F - J_{1-\tilde\omega}(z) G],\\
   (\tilde{\omega} + 1) J_{1+\tilde\omega}(z) F &= \frac{z}{2} \left[ u_0 + 
     J_{2+\tilde{\omega}}(z) F \right] + J_{1+\tilde\omega}(z) F,\\
   (-\tilde{\omega} + 1) J_{1 - \tilde\omega}(z) G &= \frac{z}{2} \left[u_0 + J_{2-\tilde{\omega}}(z) G\right] + J_{1-\tilde\omega}(z)  G.
    \end{split}
\end{equation}
Plugging the first equation into the second and third, we find
\begin{equation}
    \begin{split}
        F(\tilde{\omega},z) \left[ \left( \tilde{\omega} - \frac{z^2}{4\tilde{\omega}} \right) J_{1+\tilde{\omega}}(z) - \frac{z}{2} J_{2+\tilde{\omega}}(z) \right] &= -G(\tilde{\omega},z) \left[ \frac{z^2}{4\tilde{\omega}} J_{1-\tilde{\omega}}(z) \right],\\
        G(\tilde{\omega},z) \left[ \left( \tilde{\omega} - \frac{z^2}{4\tilde{\omega}} \right) J_{1-\tilde{\omega}}(z) + \frac{z}{2} J_{2-\tilde{\omega}}(z) \right] &= -F(\tilde{\omega},z) \left[ \frac{z^2}{4\tilde{\omega}} J_{1+\tilde{\omega}}(z) \right].
    \end{split}
\end{equation}
These equations give us a constraint on the dispersion relation:
\begin{multline}\label{disp}
	0 = \left( \tilde{\omega}^2 - \frac{z^2}{2} \right) J_{1+\tilde{\omega}}(z)J_{1-\tilde{\omega}}(z)
	- \frac{z^2}{4} J_{2+\tilde{\omega}}(z)J_{2-\tilde{\omega}}(z)\\
	+ \frac{z}{2\tilde{\omega}} \left( \tilde{\omega}^2 - \frac{z^2}{4} \right) \left[ J_{1+\tilde{\omega}}(z)J_{2-\tilde{\omega}}(z) - J_{2+\tilde{\omega}}(z)J_{1-\tilde{\omega}}(z) \right].
\end{multline}
There are infinitely many solutions $\tilde{\omega}^{(n)}(z)$ to \eqref{disp}, labelled by integers $n\in\mathbb{Z}$. For frequencies that obey \eqref{disp}, we find that the rest of the solution to \eqref{sol1} is given by
\begin{equation}\label{sol}
	\begin{split}
		G(\tilde{\omega},z) &= -F(\tilde{\omega},z) \left[ \left( \frac{4\tilde{\omega}^2}{z^2} - 1 \right) \frac{J_{1+\tilde{\omega}}(z)}{J_{1-\tilde{\omega}}(z)} - \frac{2\tilde{\omega}}{z} \frac{J_{2+\tilde{\omega}}(z)}{J_{1-\tilde{\omega}}(z)} \right],\\
		u_0(\tilde{\omega},z) &= F(\tilde{\omega},z) \left( \frac{2\tilde{\omega}}{z} J_{1+\tilde{\omega}}(z) - J_{2+\tilde{\omega}}(z) \right),
	\end{split}
\end{equation}
with the function $F(\tilde{\omega}^{(n)}(z),z)$ remaining arbitrary. Setting $F=1$ determines the `plane-wave' solution for that mode.

\subsection{Notation convention}\label{not}

From here onward we will use the following convention to label our solutions: a raised Latin index in parenthesis labels the mode/solution, while a lower Latin index labels the Fourier component of the function. For example $u^{(0)}_n$ labels the $n$th Fourier component of the zero mode. Likewise $u^{(1)}_m$ labels the $m$th Fourier component of the 1st (Goldstone) mode. The same holds for the dispersion relation; $\tilde{\omega}^{(n)}(z)$ is the dispersion relation for the $n$th mode.

\section{Spectrum of excitations and the rotons}\label{spect}

One obvious solution to the equations is
\begin{equation}
    \tilde{\omega}^{(0)}(z) = 0.
\end{equation}
We will argue, however, in Sec.~\ref{sec:Kelvin} that this mode is
unphysical.

In the low momentum limit, since $J_\alpha(z) \sim (z/2)^\alpha / \Gamma(\alpha + 1)$, the second and third term in \eqref{disp} contribute at a higher order compared to the first term, and the equation for the dispersion relation simplifies to
\begin{equation}
    \frac{z^2}{\Gamma(2+\tilde{\omega})\Gamma(2-\tilde{\omega})} \left( \tilde{\omega}^2 - \frac{z^2}{2} \right) = 0.
\end{equation}
There are infinitely many solutions to this equation, corresponding to an infinite number of branches of excitations. The lowest branch has linear dispersion relation,
\begin{equation}
	(\tilde{\omega}^{(1)})^2 = \frac{z^2}{2} \implies (\omega^{(1)})^2 = \frac{1}{2} v_F^2 q^2.
\end{equation}
This is the Nambu-Goldstone boson, whose speed is given by
\begin{equation}
    c_s^2 = \frac{v_F^2}{2} = \frac{k b}{m^2},
\end{equation}
which agrees exactly with equation (6.8) in \cite{asc}. Equation \eqref{disp} can be used to compute the next correction to the dispersion relation of the Goldstone mode, which turns out to be at order $q^3$. We find
\begin{equation}\label{sound_corr}
	\tilde{\omega}^{(1)} = \frac{z}{\sqrt{2}} \left( 1 - \frac{z^2}{64} \right).
\end{equation}

The other branches at $z=0$ correspond to the poles of the gamma functions in the denominator. Since the gamma function has poles at nonpositive integer values, these solutions correspond to (for positive frequencies only)
\begin{equation}
    \frac{\omega^{(n)}}{\omega_c} = \tilde{\omega}^{(n)} = n, \qquad n \in \mathbb{Z}, ~ n \ge 2,
\end{equation}
which are Landau levels except for $n = 0, 1$.

The equation \eqref{disp} is invariant under $\tilde{\omega} \mapsto - \tilde{\omega}$, and the spectrum is hence symmetric. The negative frequency solutions can be ignored in the usual way.

We can also compute the large momentum asymptotics of these solutions. Using the large-$z$ asymptotic expansion of the Bessel function $J_\nu(z)$, equation \eqref{disp} becomes
\begin{equation}
  - \frac{\sin\pi\tilde\omega}{4\pi\tilde\omega} z^2 - \frac{\cos2z}\pi
  + \mathcal{O}\left( \frac1z \right) = 0.
\end{equation}
which has the solutions
\begin{equation}\label{disp_as}
  \tilde\omega^{(n)} = n + \frac{(-1)^{n+1} 4n}{\pi}\,\frac{\cos2z}{z^2}
    + \mathcal{O}\left( \frac1{z^3} \right), \qquad n\in\mathbb{Z}, n \ge 0.
\end{equation}
At large momentum, the spectrum reduces to the Landau levels, but now without skipping the level with $n=1$. 
\begin{figure}[t]
    \centering
    \includegraphics[width=10cm]{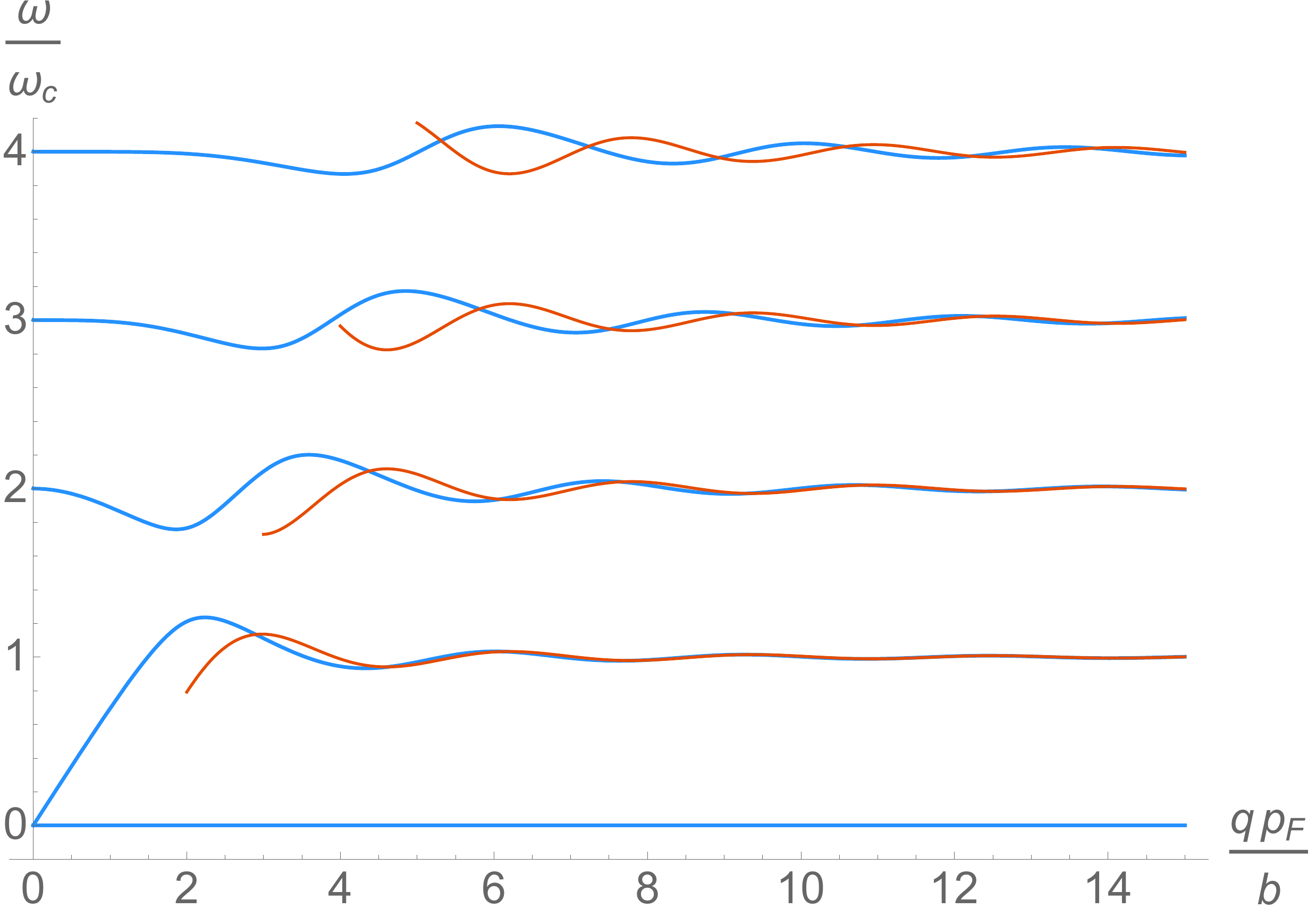}
    \caption{Spectrum of excitations for the anyon superfluid.}
    \label{displot}
\end{figure}
The numerical solutions for the first few bands are plotted in Figure
\ref{displot}. We see that the excitation spectrum of an anyon
superfluid is a deformation of Landau levels with oscillations dying out at
large momentum. The broadening of the bands also reduces for higher
Landau levels.  The minima of the dispersion curve resemble the
roton minimum in superfluid $^4$He.  Previous studies using the
random-phase approximation have found a roton minimum in the spectrum
of the Nambu-Goldstone boson in the semion gas ($k=2$ in our language)
\cite{Yang1991,Cheng1993}.

Note that the Nambu-Goldstone branch corresponds to the Landau level
index $n=1$. Here the deformation is large enough to close the gap
with the zeroth Landau level, making it the Nambu-Goldstone boson of
the superfluid. We also observe repeated magnetoroton minima (maxima)
of decreasing depth (height) as momentum increases and as we go to
higher Landau bands.

The asymptotics are plotted in red along with the dispersion relation in Figure \ref{displot}. The asymptotics are accurate in the regime $z\gg n$ for the $n$th Landau band, since this is the regime of validity of the large argument expansion of the Bessel functions. However, as evidenced by Figure \ref{displot}, the asymptotics need to be improved in the regime $z\gtrsim n$. We do so in the next section.

We do, however, have an unphysical solution with zero energy at all values of momentum. Our theory must hence be modified in order to get rid of this mode. We will return to this problem later in section \ref{kct}.

\subsection{Analysis in $\theta$-space}\label{theta_sp}

Since the asymptotics \eqref{disp_as} do not agree with the numerical solution for $z\gtrsim n$ for higher Landau bands, in this section we compute improved large momentum asymptotics for the Landau bands by recasting the equations of motion \eqref{dimless-recurs} into an integro-differential equation in $\theta$-space that takes the form:
\begin{equation}
	\begin{split}
		\tilde{\omega} u(\theta) = i\partial_\theta u(\theta) + (z \cos\theta) u (\theta) &+ i \frac{m}{b} \left( e_z e^{i\theta} + e_{\bar{z}} e^{-i\theta} \right),\\
		\frac{1}{2\pi} \int d\theta ~ e^{-i\theta} u(\theta) &= i \frac{m}{b} e_z,\\
		-\frac{1}{2\pi} \int d\theta ~ e^{i\theta} u(\theta) &= i \frac{m}{b} e_{\bar{z}}.
	\end{split}
\end{equation}
Plugging the second and third equations into the first results in
\begin{multline}
	\tilde{\omega} u(\theta) - i\partial_\theta u(\theta) - (z \cos\theta) u (\theta) = I[u,\theta]\\
	\equiv \frac{i}{\pi} \left( \sin\theta \int_0^{2\pi} d\phi ~ \cos\phi ~ u(\phi) - \cos \theta \int_0^{2\pi} d\phi ~ \sin\phi ~ u(\phi) \right).
\end{multline}
Recall that we had earlier set the momentum to be purely in the $x$-direction by invoking isotropy in order to derive this equation. Restoring the $y$-component results in the following:
\begin{equation}
	\tilde{\omega} u(\theta) - i\partial_\theta u(\theta) - \frac{p_F}{b}(q_x \cos\theta + q_y \sin\theta) u (\theta) = I[u,\theta].
\end{equation}
In what follows, it will be easier to work instead with $q_x = 0, q_y = z b / p_F$ which gives us the equation
\begin{equation}\label{ide}
	\tilde{\omega} u(\theta) - i\partial_\theta u(\theta) - (z \sin\theta) u (\theta) = I[u,\theta].
\end{equation}
We treat the momentum $z$ as a parameter, which makes this equation an eigenvalue equation for an integro-differential operator, with the eigenvalue being $\tilde{\omega}$ and the eigenfunction, $u(\theta)$.

The integro-differential equation \eqref{ide} can be written in the form of the time-independent Schr\"odinger's equation on the Hilbert space of complex functions of $\theta$ with the Hamiltonian
\begin{equation}
	H[u] = \left( i\partial_\theta + z \sin \theta \right)u(\theta) + I[u,\theta] = H_0[u] + H_1[u].
\end{equation}
In order to make this analogy precise, we need an inner product in $\theta$-space that makes the Hamiltonian Hermitian. It's easy to see that the inner product,
\begin{equation}
	\braket{f|g} \equiv \frac{1}{2\pi} \int_0^{2\pi} f^*(\theta) g(\theta) d\theta,
\end{equation}
does the trick. Suppose the solutions to this equation are given by functions $u^{(m)}(\theta)$. Since the Hamiltonian is Hermitian, and we have reduced the equations of motion to an eigenvalue problem for a Hermitian integro-differential operator, we also have an orthonormality condition on the solutions:
\begin{equation}\label{ortho}
	\braket{u^{(a)}|u^{(b)}} = \frac{1}{2\pi} \int_0^{2\pi} u^{(a)*}u^{(b)} d\theta = 0.
\end{equation}
At large $z$, the solution to $H_0$ will be rapidly oscillating and hence $H_1=I[\_,\theta]$ can be treated as a small perturbation. We can then expand the eigenvalues and eigenvectors in a perturbation series,
\begin{equation}
	\begin{split}
		u^{(a)}(\theta) &= u^{(a),0}(\theta) + u^{(a),1}(\theta) + \ldots,\\
		\tilde{\omega}^{a} &= \tilde{\omega}^{(a),0} + \tilde{\omega}^{(a),1} + \ldots.
	\end{split}
\end{equation}
The zeroth order solutions are the eigenfunctions and eigenvalues of $H_0$:
\begin{equation}
	\begin{split}
		u^{(n),0}(\theta) &= e^{-i(n\theta + z\cos\theta)},\\
		\tilde{\omega}^{(n),0} &= n \in \mathbb{Z}.
	\end{split}
\end{equation}
The correction to the `energies' at first order is then the expectation value
\begin{equation}
	\begin{split}
		\tilde{\omega}^{(n),1} &= \braket{u^{(n),0}|H_1|u^{(n),0}}\\
		&= \frac{1}{2\pi}\int_0^{2\pi} u^{(n),0*} I[u^{(n),0},\theta] ~ d\theta\\
		&= \frac{1}{(2\pi)^2} \Bigg[ \int_0^{2\pi} u^{(n),0*}(\theta) e^{i\theta} d\theta \int_0^{2\pi} u^{(n),0}(\phi) e^{-i\phi} d\phi\\
		& \qquad \qquad \qquad \qquad - \int_0^{2\pi} u^{(n),0*}(\theta) e^{-i\theta} d\theta \int_0^{2\pi} u^{(n),0}(\phi) e^{i\phi} d\phi \Bigg]\\
		&= \frac{2n}{z} J_n (z) \left( -2 \frac{\partial J_n}{\partial z} \right).
	\end{split}
\end{equation}
\begin{figure}[t]
    \centering
    \includegraphics[width=10cm]{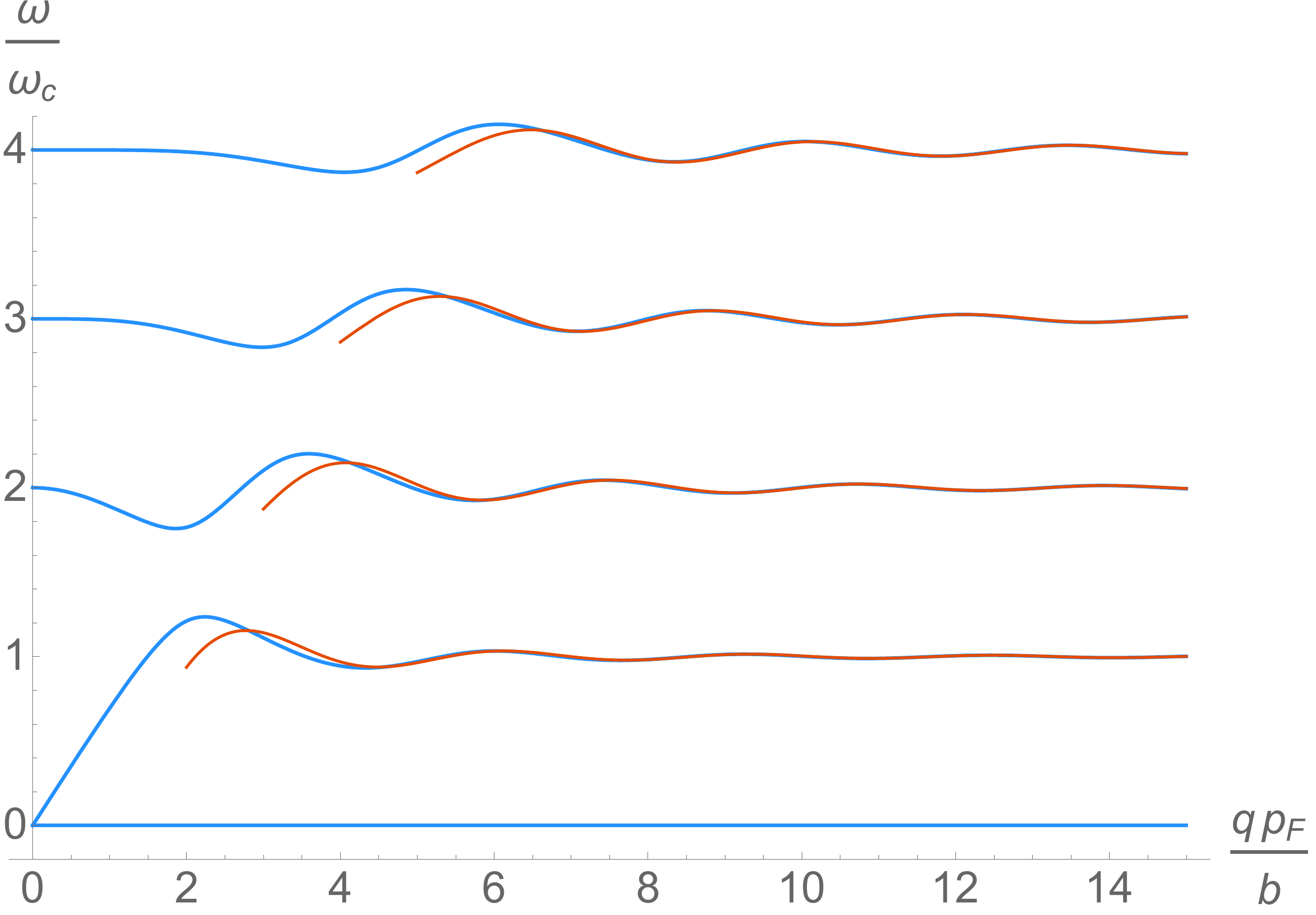}
    \caption{Improved asymptotics for the dispersion relations.}
    \label{improved_asymp}
\end{figure}
The $1/z$ factor in front of this correction verifies that $H_1$ can indeed be treated as a perturbation at large $z$. Now, if we use the asymptotic expansion for Bessel functions with large argument, we find that first order correction simplifies to
\begin{equation}
	\begin{split}
		\tilde{\omega}^{(n),1} &= \frac{2n}{z} \left[ \sqrt{\frac{2}{\pi z}} \cos \left( z - \frac{n\pi}{2} - \frac{\pi}{4} \right) \right] \left[ \sqrt{\frac{2}{\pi z}} \sin \left( z - \frac{n\pi}{2} - \frac{\pi}{4} \right) \right]\\
		&= \frac{(-1)^{n+1}4n}{\pi} \frac{\cos 2z}{z^2},
	\end{split}
\end{equation}
which agrees with \eqref{disp_as}. We can also compute more accurate asymptotics by using the large order, large argument asymptotic form of the Bessel functions:
\begin{equation}
	J_\nu (\nu \sec\beta) = \sqrt{\frac{2}{\pi \nu \tan \beta}} \cos \left[ \nu (\tan\beta - \beta) - \frac{\pi}{4} \right] + \mathcal{O}\left( \frac{1}{\nu} \right),
\end{equation}
with $\nu = n$ and $\beta = \cos^{-1}(n/z)$. This expression is only valid in the $n$th Landau level for the region $z\ge n$. Working out the leading order behaviour of the dispersion, we find
\begin{equation}
	\tilde{\omega}^{(n),1} = n - \frac{4n}{\pi z^2} \cos \left[ 2\sqrt{n^2 - z^2} - 2n \cos^{-1}\frac{n}{z} \right].
\end{equation}
Note that for $z \gg n$ we recover the large $z$ asymptotics of \eqref{disp_as}. The improved asymptotics are plotted in Figure \ref{improved_asymp}. The improved asymptotic form is fairly accurate beyond the second minimum in each Landau level.

\section{Kelvin's circulation theorem and the unphysical nature of the zero mode}\label{kct}
\label{sec:Kelvin}

We now return to the problem of the mode with zero energy.  We will
argue in this Section that this mode is unphysical and has to be
eliminated from the quantum theory.  The idea is that such a mode
corresponds to an operator (more precisely, an infinite number of
operators) which commutes with every other operator in the algebra,
hence the states where this operator has nonzero expectation value are
unphysical, since they are inaccessible from the ground state by any
unitary transformation.

The operator that we will construct is a generalization of the
vorticity in ideal hydrodynamics, which gives rise to the Kelvin
circulation theorem.  In a generalization of ideal hydrodynamics
called ``chiral metric hydrodynamics''~\cite{cmh}, which includes
spin-2 operators in addition to the density and velocity, such an
operator also exists.  Inspired by these examples, we search for an
operator of the form
\begin{equation}
	\tilde\Omega (q) = a_0 (q^2) \bar{u}_0 + \sum_{n=1}^\infty \frac{a_n(q^2)}{(q/2)^n} \left[ q_{\bar{z}}^n \bar{u}_n + (-1)^n q_z^n \bar{u}_{-n} \right]
\end{equation}
Requiring this operator to commute with all other operators in the theory
leads to the following recursion relation
\begin{equation}
	z(a_{n+1} + a_{n-1}) - 2 n a_n = 0, \qquad n\ge 1
\end{equation}
Demanding that $a_n\to0$ when $n\to\infty$, the solution to this recursion
relation is, up to an overall coefficient
\begin{equation}
  a_n (q^2) = J_n (z)
\end{equation}
so
\begin{equation}\label{cas_bar}
  \tilde\Omega(q) =  J_0(z) \bar{u}_0 + \sum_{n=1}^\infty \frac{J_n(z)}{(q/2)^n} \left( q_{\bar{z}}^n \bar{u}_n + (-1)^n q_z^n \bar{u}_{-n} \right)
\end{equation}
This can be written in terms of the ``unbarred'' $u_n$ fields.  By
using $2 (q_z \delta a_{\bar{z}} - q_{\bar{z}} \delta a_z) = \delta b
= p_F u_0 / k$ we find
\begin{equation}\label{cas_unbar}
   \tilde\Omega(q) =  -J_2(z) u_0 + \sum_{n=1}^\infty \frac{J_n(z)}{(q/2)^n} \left( q_{\bar{z}}^n u_n + (-1)^n q_z^n u_{-n} \right)
\end{equation}

Let us now show that $\tilde\Omega(q)$ is nonzero only on the mode
with zero energy.  Setting $q_x = q, q_y = 0$ without loss of
generality in \eqref{cas_unbar}, we find
\begin{equation}
	\begin{split}
		\frac{\Omega}{n}(q) &= - J_2(z) u_0 (q) + \sum_{n \ge 1} \left[ J_n(z) u_n(q) + (-1)^n J_n u_{-n}(q) \right]\\
		&= -J_2 (z) u_0 (q) + \sum_{n\ge 1} J_n(z) u_n(q) + \sum_{n\le -1} (-1)^n J_{-n} (z) u_n(q)
	\end{split}
\end{equation}
Now the solution for the zero mode is given by plugging in $\tilde{\omega}=0$ into \eqref{sol1} and \eqref{sol}.  Up to an overall factor,
\begin{equation}\label{zero}
	\begin{split}
		u^{(0)}_0(z) &= - J_2(z)\\
		u^{(0)}_n(z) &= \begin{cases} J_{n}(z), \qquad &n \ge 1\\
		(-1)^n J_{-n}(z), \qquad &n \le -1
		\end{cases}
	\end{split}
\end{equation}
Using \eqref{zero}, the value of the Casimir, evaluated on a mode with
amplitudes $u_n(q)$, can be written as
\begin{equation}
  \frac{\Omega}{z}(q) = \sum_{n=-\infty}^{\infty} u^{(0)}_n(q) u_n(q) = \sum_{n=-\infty}^{\infty} u^{(0)*}_n(q) u_n(q)
\end{equation}
Using the $\theta$-space formalism of section \ref{theta_sp}, we see
that the Casimir can also be written as the inner product
\begin{equation}
   \frac{\Omega}{n}[u] = \int_0^{2\pi} \frac{d\theta}{2\pi} u^{(0)*}(\theta) u(\theta) = \braket{u^{(0)}|u}
\end{equation} 
In particular, orthogonality of the solutions then implies that
\begin{equation}
  \frac{\Omega}{n}[u^{(m)}(\theta)] = \braket{u^{(0)}|u^{(m)}} \propto \delta^{0m}
\end{equation}
Hence the Casimir vanishes for the non-zero modes, and does not vanish
for the zero mode.  Thus the mode with zero frequency is unphysical
and should be eliminated from the spectrum.

\section{Conslusion and Discussion}

In this paper, we have calculated the excitation spectrum of a gas of
nonrelativistic anyons with statistical angle $\theta=\pi(1-\frac 1k)$
for integer and large $k$.  We have shown that the spectrum consists
of discrete branches which, at large momenta, have energy given by the
energy levels of the Landau levels in the effective magnetic field.
At small momenta these ``Landau levels'' are distorted, and the lowest
excitation branch turns into a Nambu-Goldstone boson.  Each branch of
the spectrum possesses series of roton maxima and minima.

That at large momentum the energy levels look like Landau levels can
be explained as follows.  The excitations considered in this paper are
neutral excitations and can be interpreted as electron-hole bound
states.  In a magnetic field the distance between the particle and the
hole is proportional to the momentum carried by the pair, thus at
large momentum the distance between the two particles are large.
The interaction between the particle and the hole can be neglected,
and as both these particles live in a nonzero average magnetic field,
is then natural that the energy levels become multiples of the
cyclotron frequency.

It is interesting to compare our result with the spectrum of
magnetorotons in the fractional quantum Hall effect~\cite{fqh}. While our model is not directly applicable to fractional quantum Hall states, a very similar procedure can be used to compute the dispersion relations of excitations in FQH states, as in \cite{fqh}. These
excitations also have roton-like minima and maxima.  Our picture implies
then that the dispersion curves of the neutral excitations in the FQHE
also look like Landau levels at large momentum.  This is consistent
with the magnetorotons being a well separated particle-hole pair at
large momentum.

The calculations performed in this paper are done to the leading
order in the expansion over $1/k$.  To this order the excitations
that we found are stable.  To find the decay rates of the higher
excitations one needs to go to subleading orders in $1/k$. These also introduce nonlinearities in the Poisson bracket \eqref{pb} as well as the Hamiltonian \eqref{ham} and it is unclear apriori whether this expansion is resummable. We defer these calculations to future work.

We thank Alexander Bogatskiy for discussion.  This work is supported, in
part, by the U.S.\ DOE grant No.\ DE-FG02-13ER41958, a Simons
Investigator grant and by the Simons Collaboration on Ultra-Quantum
Matter, which is a grant from the Simons Foundation (651440, DTS).

\bibliographystyle{JHEP}
\bibliography{AnyonSuperfluids}

\end{document}